\documentstyle[12pt]{article}

\evensidemargin=\oddsidemargin
\begin{document}
\begin{flushright}
{\large \sf  VPI-IHEP-93-13\\}
\end{flushright}
\bigskip
\bigskip
\begin{center}
{\large \bf Inconsistency of QED in the Presence of Dirac Monopoles II.
\footnote{Work supported in part by the U.S. Department of Energy under
           grant DEFG05-92ER40749.} }\\[1.5cm]

{\bf Hong-Jian He }$^{a,b}$ ~~~~{\bf Zhaoming Qiu }$^b$
{}~~~~{\bf Chia-Hsiung Tze }$^a$  \\[0.7cm]

$^a$ Department of Physics and Institute for High Energy Physics\\
     Virginia Polytechnic Institute and State University\\
     Blacksburg, Virginia 24061-0435, U.S.A.
      \footnote{Present address of H.J. He.}\\

$^b$ China Center of Advanced Science and Technology (World Laboratory) \\
     P. O. Box 8730, Beijing 100080, P.R. China
\end{center}

\bigskip\bigskip
\centerline{\bf Abstract}
\bigskip
\begin{center}
\begin{minipage}[c]{15cm}
\begin{sf}
\parindent=1.2cm

\noindent
We enlarge the local gauge invariance of QED from $~U(1)_A~$ to
$~U(1)_A \times U(1)_{\Theta}~$ by introducing another unphysical
pure gauge field $~\Theta~$ with an independent, unphysical gauge coupling
$~\tilde{e}~$. This pure gauge field can be gauge-transformed away and
the resulting theory is {\it identical to} standard QED.
We then re-examine the Dirac quantization condition (DQC) for point
monopoles and find that two essentially different DQCs can be derived.
One DQC involves a
gauge coupling $~e~$ in the $~U(1)_A~$ group and the other
only the unphysical gauge coupling $~\tilde{e}~$ in the
$~U(1)_{\Theta}~$ group. The unique physically consistent solution of these
two DQCs is a vanishing magnetic charge, which implies that no Dirac
monopole exists in nature.

\vspace{2.5cm}
\noindent
\end{sf}
\end{minipage}
\end{center}

\newpage
\begin{sf}

In this second of a series of papers$^{[1]}$ devoted to point monopoles in
QED, we present an alternative proof of the
inconsistencies of QED in the presence of Dirac monopoles. We noticed that
the singular Dirac string$^{[2]}$ in the monopole gauge potential is purely
a gauge-artifact. It is just the gauge freedom which
allows us to arbitrarily move the string around without any physical effect,
provided that a consistent condition
--- Dirac quantization condition (DQC)$^{[2]}$ is
satisfied. By introducing another unphysical pure gauge field into QED, we
find it possible to attribute part of the singularities to this pure
gauge field and thus the corresponding DQC involves the unphysical
gauge coupling associated with this pure gauge field. So the physically
consistent solution to both the original DQC and this new DQC can only
be a vanishing magnetic charge. In Sec.1, a generalized QED Lagrangian with
an enlarged local gauge symmetry $~U(1)_A\times U(1)_{\Theta}~$ is proved
to be identical to standard QED up to the quantum-field-theory-level. Of
course,
the gauge coupling associated with this pure gauge field in the
$~U(1)_{\Theta}~$ group is shown to be entirely arbitrary. Two independent
DQCs are carefully derived in Sec.2 and some conclusions are  given in Sec.3.

\noindent
{\bf 1. A generalized QED Lagrangian and the Ward-Takahashi identities}

An Abelian or non-Abelian  global symmetry can always be localized by
introducing an {\it unphysical pure gauge field}, which has no kinetic term
and can be gauge-transformed away.
A dynamical gauge field is only a natural generalization
and at present its existence can be determined only by experiments.
A pure gauge field is {\it sufficient and necessary }
to insure the ordinary local gauge invariance.
This may be why without discovering
the corresponding dynamical gauge fields
we have observed a lot of global symmetries (such as
the lepton and baryon numbers conservations) which had been independently
tested at different local places.
Besides the electric charge conservation, standard QED
has an extra global $~U(1)~$ symmetry which is the electron number
conservation. In the following we shall localize this extra
global $~U(1)~$ symmetry by introducing a pure gauge field.
One should notice that only the physical gauge coupling
associated with a dynamical gauge field can be related to its global charge
and the unphysical gauge coupling associated with the pure gauge field has
nothing to do with the global charge since it is non-observable and the
corresponding pure gauge field can be completely  gauge-transformed away.

But when including monopoles, we should carefully distinguish
two essentially different situations.
In the Dirac monopole case, a singular gauge transformation must be allowed
in order to arbitrarily move the Dirac string and thus make it non-observable
as desired. This singular $~U(1)~$ gauge transformation
( which is usually called as an "extended" gauge transformation$^{[3]}$ ) can
thus arbitrarily  change the pure gradient part of the monopole's gauge field
or even entirely transform it away while leaving the physical magnetic field
invariant. This is in sharp contrast with the case of the spatially extended
't Hooft-Polyakov monopole$^{[4]}$ which, as finite energy solution to the
spontaneously broken gauge theories, is naturally singularity-free at
the beginning. All allowed regular gauge transformations cannot rid of
the pure gauge field (or even change their homotopy class). Furthermore
any singular gauge transformation which transforms the pure gauge field away
must be forbidden since it leaves a vanishing magnetic field.

In this section we first discuss QED without Dirac monopoles.
Consider the following generalized QED Lagrangian
\begin{equation}
{\cal L}=-\frac{1}{4}F_{\mu\nu}F^{\mu\nu}+\bar{\psi} i \gamma_\mu
D^\mu\psi-m\bar{\psi}\psi                                     
\end{equation}
with
\begin{equation}
\begin{array}{l}
D_\mu = (\partial_\mu-i e A_\mu
-i{\tilde e}\partial_\mu\Theta ) ~, \\
F_{\mu\nu} = \partial_{\mu}\bar{A}_{\nu}-\partial_{\nu}\bar{A}_{\mu}
           = \partial_{\mu}A_\nu-\partial_{\nu}A_{\mu} ~, \\
\bar{A}_{\mu}\equiv A_{\mu} + \partial_{\mu}\Theta ~,\\
\end{array}
\end{equation}                                               
where  $~\bar{A}_{\mu}~$ is only a notation
in which the coefficient of $~\partial_{\mu}\Theta~$ is arbitrary
but can always be chosen to be unity
since the {\it unphysical} $~\Theta~$ field
has no kinetic term and can be arbitrarily rescaled without any physical
effect.
The above QED Lagrangian has a larger
local symmetry $~U(1)_A\times U(1)_{\Theta}~$, i.e. it is invariant under
the following two kinds of independent gauge transformations:
\begin{description}
\item[(i).]      
The $~U(1)_A~$ gauge transformation
\begin{equation}
\begin{array}{l}
\psi '(x)= e^{-i\zeta (x)}\psi (x) ~,~~~
\bar{\psi}'(x) = \bar{\psi}(x)e^{i\zeta (x)}~;~\\
A_{\mu}' = A_{\mu} - e^{-1}\partial_{\mu}\zeta (x) ~,\\
\Theta ' = \Theta ~.
\end{array}
\end{equation}                                    

\item[(ii).]     
The $~U(1)_{\Theta}~$ gauge transformation
\begin{equation}
\begin{array}{l}
\psi '(x)= e^{-i\eta (x)}\psi (x) ~,~~~
\bar{\psi}' (x) = \bar{\psi}(x)e^{i\eta (x)} ~; \\
A_{\mu}' = A_{\mu} ~,\\
\Theta ' = \Theta - \tilde{e}^{-1}\eta (x) ~.
\end{array}
\end{equation}                                  
\end{description}
Eq.(4) clearly shows that the unphysical pure gauge field can be
completely gauge-transformed away and thus our generalized QED
Lagrangian simply reduces to the standard QED Lagrangian.
Actually the standard QED is in the {\it "unitary gauge"} of eq.(1),
in which the pure gauge field has been transformed away.
Here it is clear that the gauge coupling $~e~$ and $~\tilde{e}~$ belong
to the two direct product group $~U(1)_A~$ and $~U(1)_{\Theta}~$ respectively,
and thus are {\it independent of each other}.

{}From (1) and (2), the definition of the magnetic field is
\begin{equation}
\vec{B} = \vec{\bigtriangledown}\times\vec{\bar{A}}
        = \vec{\bigtriangledown}\times\vec{A} ~.
\end{equation}                                  

The nonintegrable phase factor is now expressed as
\begin{equation}
P(x_2, x_1; C) = \exp\left[ ie\int^{x_2}_{x_1} A_{\mu}dx^{\mu}
    + i\tilde{e}\int^{x_2}_{x_1}
      \partial_{\mu}{\Theta}dx^{\mu}\right] ~.
\end{equation}                                   

When doing quantization, we need two gauge-fixing terms for two gauge
groups $~U(1)_A~$ and $~U(1)_{\Theta}~$ respectively, i.e.
\begin{equation}
L_{gf} = -\frac{1}{2\xi_{A}}F_1(A)^2
         -\frac{1}{2\xi_{\Theta}}F_2(\Theta)^2 ~.
\end{equation}                                   
For example, the gauge-fixing functions $~F_1(A)~$ and $~F_2(\Theta )~$
can be chosen as
\begin{equation}
F_1(A)=\partial^{\mu}A_{\mu} ~,~~~~ F_2(\Theta ) = \partial^2\Theta ~.
\end{equation}                                    
Now we derive some new Ward-Takahashi (WT) identities
for the $~U(1)_{\Theta}~$ gauge group. By introducing the external sources
$~~~J_{\mu}A^{\mu} + \bar{I}\psi + \bar{\psi}I~~~$ in the generating
functional for Green functions and doing a $~U(1)_{\Theta}~$ gauge
transformation, we can easily
re-derive the following generating equation
\begin{equation}                                  
(\xi_{\Theta}\tilde{e})^{-1}\partial^4\Theta + \tilde{e}^{-1}
\frac{\delta\Gamma}{\delta\Theta} =
i\left[\frac{\delta\Gamma }{\delta\psi }\psi
+\bar{\psi}\frac{\delta\Gamma}{\delta\bar{\psi}}\right] ~.
\end{equation}
{}From (9) we get the following two WT identities
\begin{equation}                                        
\begin{array}{l}
i\tilde{D}^{-1}(k) = -\xi^{-1}_{\Theta}k^4 ~,\\
(p_{\mu}'-p_{\mu})\Lambda^{\mu}(p',p) = iS^{-1}(p')-iS^{-1}(p) ~,
\end{array}
\end{equation}
where
\begin{equation}                                        
i\tilde{D}^{-1}(k) = \int_{FT}
\delta^2\Gamma /[\delta
\Theta (y) \delta \Theta (x)] ,~~~
\tilde{e}(p_{\mu}'-p_{\mu})\Lambda_{\mu}(p',p)
\equiv \int_{FT} \delta^3\Gamma /
[\delta\psi(z)\delta\bar{\psi}(y) \delta \Theta (x)] ,
\end{equation}
($\int_{FT}$ denotes the Fourier transform)
and $S(p)$ is the full fermion propagator. Also we can easily find that
\begin{equation}                                        
\int_{FT} \delta^3\Gamma / [\delta
\psi(z)\delta\bar{\psi}(y) \delta A^{\mu}(x)]
= e\Lambda_{\mu}(p',p)  .
\end{equation}

To perform the renormalization, we define
\begin{equation}                                          
\begin{array}{l}
\psi=Z_2^{\frac{1}{2}}\psi_R,~~~
\bar{\psi}=Z_2^{\frac{1}{2}}\bar{\psi}_R,~~~
A^{\mu} = Z_A^{\frac{1}{2}}A^{\mu}_R,~~~
\Theta = Z_{\Theta}^{\frac{1}{2}}\Theta_R,\\
m=Z_mm_R,~~~ e=Z_ee_R,~~~ \tilde{e}= Z_{\tilde{e}}\tilde{e}_R,~~~
 \xi_A = Z_{\xi_A}\xi_{AR},~~~ \xi_{\Theta}=Z_{\xi_{\Theta}}\xi_{\Theta R}~.
\end{array}
\end{equation}
Here the non-observable gauge coupling $~\tilde{e}~$ of the
pure gauge field $~\Theta~$ has an independent renormalization
constant $~Z_{\tilde{e}}~$.

We rewrite (1) as
\begin{equation}                                           
{\cal L}=Z_A \frac{-1}{4}F_{R\mu\nu}F_R^{\mu\nu} +
Z_2\bar{\psi}_R (i{\not\partial} - Z_mm_R){\psi}_R +
Z_1e_R A_R^{\mu}\bar{\psi}_R\gamma_{\mu}\psi_R +
\tilde{Z}_1\tilde{e}_R\partial^{\mu}\Theta_R\bar{\psi}_R\gamma_{\mu}\psi_R.
\end{equation}
Then we have
\begin{equation}
Z_e = Z_1 Z_2^{-1} Z_A^{-\frac{1}{2}},~~~~~~
Z_{\tilde{e}} = \tilde{Z_1} Z_2^{-1} Z_{\Theta}^{-\frac{1}{2}}.
\end{equation}
The WT identity (10) {\it only requires} that, after renormalization,
$~~~~Z_{\xi_{\Theta}} = Z_{\Theta} ~~~~$,
where {\it either $~Z_{\xi}~$ or $~\tilde{Z}_A~$ but not both
can be arbitrarily chosen}.
Since (10) shows that $\tilde{D}_{\mu\nu}$ has no loop correction at all,
the most natural and simplest choice is
\begin{equation}
Z_{\xi_{\Theta}} = Z_{\Theta} = 1 ~.
\end{equation}
In general, we can choose
$~~~Z_{\xi_{\Theta}}=Z_{\Theta}=1 + (~arbitrary~loop-order~quantities~)
{}~~~$.    The WT identity (11)
and eqs.(12)(13) give $~~~Z_1 = \tilde{Z}_1 = Z_2~~~$.
So substituting this equation and (16) into (15) we get
\begin{equation}
Z_e = Z_A^{-\frac{1}{2}} ,~~~~~~~
Z_{\tilde{e}} = Z_{\xi_{\Theta}}^{-\frac{1}{2}} = 1 .
\end{equation}
In consequence we prove that the renormalization for $\tilde{e}$ is actually
arbitrary and may need no renormalization whatsoever. This is not surprising
since for the product groups $~U(1)_A\times U(1)_{\Theta}~$, the gauge
coupling $~\tilde{e}~$ of $~U(1)_{\Theta}~$ has nothing to do with the
the physical coupling $~e~$ of $~U(1)_A~$.

Finally, we emphasize again
that our above generalized QED is {\it identical to} standard QED,
even up to loop-level. Clearly, the introduction of a pure gauge
field which can be gauge-transformed away can have no physical effects.

\noindent
{\bf 2. Dirac quantization condition re-examined }

Following our part-I we still work in the standard Dirac
formulation$^{[5]}$.
Let us consider a Dirac monopole $g$ with magnetic field
$~~~~ \vec{B}(x)=
\displaystyle\frac{g}{r^2}\frac{\vec{r}}{r} ~~~$,
where $ r = \mid\vec{x}\mid $.
The magnetic field is related to the
monopole's gauge potential by
$~~\vec{B} = \vec{\bigtriangledown}\times \vec{\bar{A}}~$ which implies that
$~\bar{A}_{\mu}~$ cannot be regular everywhere and must contain some
singularities.  Since the physical $~\vec{B}~$ field is regular
everywhere except at the origin,
in the standard Dirac formulation$^{[3]}$, the above definition
is modified by adding the so-called Dirac string to cancel the
singularities in $~\vec{\bigtriangledown}\times \vec{\bar{A}}~$, so that
the correct $~\vec{B}~$ field is obtained.
Following the same steps as before, we obtain the two simplest Dirac
solutions for $~~\bar{A}_{\mu} (\equiv A_{\mu} +\partial_{\mu}\Theta )~~$
with singular lines along the negative and positive $z$-axes,
respectively:
\begin{equation}                                         
(\bar{A}_{\mp\hat{z}})_t=(\bar{A}_{\mp\hat{z}})_{r}
=(\bar{A}_{\mp\hat{z}})_{\theta}=0~,~~(\bar{A}_{\mp\hat{z}})_{\varphi}
=\displaystyle\frac{g}{r}\frac{\pm 1-\cos\theta}{\sin\theta} ~.
\end{equation}
They are connected by the gauge transformation
\begin{equation}                                          
\bar{A}^{\mu}_{\hat{z}}= \bar{A}^{\mu}_{-\hat{z}}
                           -\partial^{\mu}(2g\varphi )~.
\end{equation}
{}From (3) and (4), we see that this can be regarded as a gauge transformation
of $~U(1)_{A}~$ with
\begin{equation}                                          
\begin{array}{l}
A^{\mu}_{\hat{z}} = A^{\mu}_{-\hat{z}}- e^{-1}\partial^{\mu}\zeta (x)~,~~~~
\zeta = 2e g\varphi ~~\\
\vec{A}_{\hat{z}}
=\displaystyle\frac{g}{r}\displaystyle\frac{1-\cos\theta}{\sin\theta}
= \vec{A}_{-\hat{z}}-\displaystyle\frac{2g}{r\sin\theta}\hat{\varphi}~~,\\
\Theta_{\hat{z}} = \Theta_{-\hat{z}} =0 ~~;
\end{array}
\end{equation}
{\it or}, a gauge transformation of $~U(1)_{\Theta}~$ with
\begin{equation}                                           
\begin{array}{l}
\vec{A}_{\hat{z}}= \vec{A}_{-\hat{z}}
=\displaystyle\frac{-g\cos\theta}{r\sin\theta}\hat{\varphi} ~~,\\
\Theta_{\hat{z}}=\Theta_{-\hat{z}} - \tilde{e}^{-1}\eta (x)~,~~~~
\eta = 2\tilde{e}g\varphi~~,\\
\Theta_{\hat{z}}= -g\varphi = -\Theta_{-\hat{z}} ~~.
\end{array}
\end{equation}

Now we can repeat the three standard approaches given in part-I to derive
the DQC by using the above two kinds of gauge potentials
and their transformations in (20) and (21), respectively.
Thus, from (20) we just obtain the ordinary DQC
\begin{equation}                                  
e g =\frac{n}{2} ~, ~~(n=0, \pm 1, \pm 2, \cdots )~~;
\end{equation}
while from (21) we get an {\it independent new DQC}
\begin{equation}                                  
\tilde{e} g =\frac{k}{2} ~,~~(k=0,\pm 1, \pm 2, \cdots )~~,
\end{equation}
which has a similar form to (22) but has a completely
different physical meaning.
Here an important observation is that in (23)
{\it the physical magnetic charge $g$
is constrained by the non-observable gauge coupling $\tilde{e}$
for $k\neq 0$}. This is not surprising
since the {\it singular Dirac string is a pure gauge artifact and thus
can be naturally attributed to an unphysical pure gaue field.}
It is easy to check that for the case of the 't Hooft-Polyakov monopole$^{[4]}$
the DQC (23) cannot be derived even if one introduces an extra unphysical
$~U(1)~$ pure gauge field, since there is no singularity. Also
the original consistent condition (22) is unnecessary and the electric
charge is automatically quantized in the 't Hooft-Polyakov monopole case.

\noindent
 {\bf 3. Inconsistency of QED in the presence of Dirac monopoles}

In (22) and (23) the gauge couplings $~e~$ and $~\tilde{e}~$ belong to
two direct product $~U(1)~$ groups respectively
and thus are {\it independent of each
other} as we pointed out before.
There are actually {\it two possible} solutions to the original
DQC (22): $~~g=0~~$ with
$~~n=0~~$ and $g\neq 0$ with $~~n\neq 0~~$.  However, in our new
DQC (23) the only physically consistent solution is
$~~~g=0~~~$ with $~k=0~$, which is also a possible solution to DQC (22).
The nonvanishing solution $~~g\neq 0~~$ in (23) constrains the
physical magnetic charge $~g~$ with unphysical coupling $~\tilde{e}~$
and thus can never be
consistent as already analyzed in our part-I. Hence we conclude that
the unique physically reasonable solution to both (22) and (23) is
$~~g=0~~$, which implies that {\it no Dirac monopoles exist in the nature}.
Thus this alternative proof strengthens our conclusion in part-I
from a different point of view.
Other inconsistencies of Dirac monopoles are presented
elsewhere$^{[6]}$.

\null
\noindent
{\bf Acknowledgement }
 H.J. He thanks Prof. Lay Nam Chang for discussions.

\noindent
{\bf References }
\begin{enumerate}
\item                
H.J. He, Z. Qiu, C.H. Tze, VPI-IHEP-93-12;

\item                 
P.A.M. Dirac, Proc. Roy. Soc. {\bf A133}(1931)60. An extension of this
idea to include relativity is given in his another paper: Phys.Rev.
{\bf 74}(1940)817. After Dirac's 1st paper, thousands of papers on
this subject have since appeared. For a more detailed list,
see, for example, {\it Resource Letter MM-1: Magnetic Monopoles}
by A.S. Goldhaber, W.P. Trower in Am.J. Phys. {\bf 58}, 429(1990).

\item                 
See, for example,
P. Goddard and  D.I. Olive, Rep.Prog.Phys. {\bf 41}(1978)1357.

\item  
G. 't Hooft, Nucl. Phys. {\bf B79}, 276(1974);\\
A.M. Polyakov, Sov. Phys. JETP Lett. {\bf 20}, 194(1974).

\item                  
For general reviews, see for examples: B. Zumino, "{\it Recent developments
in the theory of magnetically charged particles}" in {\it Strong and Weak
Interactions---Present Problems},
({\it Proc. 1966 Int. School of Physics 'Ettore Majorana'} ), p.711,
ed. A. Zichichi, Academic Press, New York and London;
P. Goddard and  D.I. Olive, in Ref.[3].

\item  
H.J. He, Z. Qiu, C.H. Tze, VPI-IHEP-93-14;~\\
H.J. He, C.H. Tze, VPI-IHEP-93-08, 09, 10, 11.
\end{enumerate}

\end{sf}
\end{document}